\documentclass[amsmath,amsfonts,aps,nofootinbib,preprintnumbers]{revtex4}
\usepackage{graphicx}
\usepackage{amssymb}
\usepackage{epsfig}
\begin{document}
\baselineskip=20pt
\title{Ground-state energy shift of $n$ pions and $m$ kaons in a finite volume}
\author{Brian Smigielski}\email{smigs@u.washington.edu} \affiliation{Department of Physics,
  University of Washington\\ Box 351560, Seattle, WA 98195, USA}
\author{Joseph Wasem}\email{wasem@u.washington.edu} \affiliation{Department of Physics,
  University of Washington\\ Box 351560, Seattle, WA 98195, USA}

\preprint{NT@UW-08-25}

\begin{abstract}
The ground state energy of a collection of $n$ pions and $m$ kaons
with short range interactions is calculated for a volume with finite
spatial extent $L$ and periodic boundary conditions. This
calculation is accomplished to order $L^{-6}$ in the large volume
expansion. With this result one can extract the various two- and
three-body interactions between pions and kaons from lattice QCD
data.
\end{abstract}

\maketitle

Recently, efforts have been made using lattice QCD to extract both
the $\pi^{+}\pi^{+}$ and the $K^{+}K^{+}$ scattering lengths as well
as the three-body interactions $\pi^{+}\pi^{+}\pi^{+}$ and
$K^{+}K^{+}K^{+}$\cite{Beane:2007es,Detmold:2008fn,Detmold:2008yn,Detmold:2008bp}.
These efforts have made steps toward using lattice QCD to rigorously
determine the nuclear equation of state (NEOS) directly from the
underlying theory of QCD. A knowledge of the NEOS would be of great
importance to many research areas, including helping to predict the
evolution of supernovae. In addition, an understanding of multi-pion
and multi-kaon systems provide insight into strongly interacting
boson gases. References
\cite{Beane:2007es,Detmold:2008fn,Detmold:2008yn,Detmold:2008bp}
have made use of a method involving the volume dependence of the
energy spectrum (below inelastic thresholds) of two hadrons as a
function of their scattering
length\cite{Luscher:1986pf,Luscher:1990ux}, where lattice QCD is
used to calculate the energy of multi-hadron states and this result
is then applied to the calculated dependence of the finite volume
ground state energy shift to determine interaction strengths. The
ground state energy of \textit{n} identical bosons with short range
interactions is calculated in the large volume expansion to
$\mathcal{O}(L^{-7})$ in Ref. \cite{Detmold:2008gh}, providing a way
to extract the three-body interaction strength between identical
bosons which enter at $\mathcal{O}(L^{-6})$, and building upon the
earlier work of Refs.
\cite{Beane:2007qr,Huang:1957im,Wu:1959zz,Hugenholtz:1959zz,Sawada:1959zz}.
In this work we extend the calculation to determine the ground state
energy shift in finite volume of \textit{n} pions and \textit{m}
kaons to $\mathcal{O}(L^{-6})$ in the large volume expansion using a
multispecies extension of the techniques used in Refs.
\cite{Beane:2007qr,Detmold:2008gh}. This result will allow for the
systematic extraction of $\pi K$ scattering lengths and the
three-body interactions $\pi\pi{K}$ and $\pi{KK}$, as well as
provide a different way to extract $\pi\pi$ and $KK$ scattering
lengths by using mixed pion-kaon systems in lattice calculations.

To calculate the ground state energy there are multiple (and
entirely equivalent) methods that one can use. One such method uses
a resolvent function of the system Hamiltonian defined
by\cite{Luscher:1986pf}:
\begin{eqnarray}\label{resolvent}
    F(z)&=&\langle0|\frac{1}{z-\mathcal{H}}|0\rangle\nonumber\\
    &=&\frac{1}{z-E_{0}-r(z)}
\end{eqnarray}
with
\begin{equation}
r(z) = \langle 0| \hat{V} \sum_{n=0}^{\infty} \left(
\frac{\hat{Q}_0}{z-\mathcal{H}_0} \hat{V} \right)^n|0 \rangle
\end{equation}
where $|0 \rangle$ is the ground-state of the system and
$\hat{Q}_{0}=1-|0\rangle\langle0|$ is a projection operator
discussed below. Here $\mathcal{H}$ is the full Hamiltonian of the
system while $\mathcal{H}_{0}$ is the free Hamiltonian. In eqn.
(\ref{resolvent}), when $F(z)$ takes the form found in the first
line, a perturbative expansion leads to a pole near $z=E_0$.
However, projecting out that portion of the propagator containing
the pole using the operator $\hat{Q}_{0}$ leads to the function
$r(z)$ in the second line which is smooth in the neighborhood of
$E_0$ and which can be perturbatively expanded in powers of the
potential $\hat{V}$. To find the ground state energy shift one needs
to expand the function $r(z)$ around the point $z=E_0$. The leading
terms in this expansion are\cite{Luscher:1986pf}:
\begin{eqnarray}
    r_{j}(z)&=&\frac{1}{j!}\frac{\partial^{j}}{\partial{z}^{j}}r(z)|_{z=E_0}\nonumber\\
    \Delta{E}&=&r_{0}(E_0)+r_{0}(E_0)r_{1}(E_0)+r_{0}(E_0)r_{1}^2(E_0)+r_{0}^2(E_0)r_{2}(E_0)+... \ .
\end{eqnarray}

The method of pseudo-potentials can be employed to determine $r(z)$
above, or one can use the different but equivalent approach of
nonrelativistic time-independent perturbation theory (NRPT). The
potential used in this calculation is:
\begin{align}\label{potential}
\hat{V} & =
\sum_{i<j\in\mathcal{P}_{\pi}}\chi_{\pi}\delta^{(3)}(\vec{r}_i -
\vec{r}_j) +
\sum_{i\in\mathcal{P}_{\pi\pi},j\in\mathcal{P}_{K}}\chi_{\pi
K}\delta^{(3)}(\vec{r}_i - \vec{r}_j)
+  \sum_{i<j\in\mathcal{P}_{K}}\chi_{KK}\delta^{(3)}(\vec{r}_i - \vec{r}_j) + \nonumber \\
&
\sum_{i<j<k\in\mathcal{P}_{\pi}}\eta_{3,\pi\pi\pi}\delta^{(3)}(\vec{r}_i
- \vec{r}_j)\delta^{(3)}(\vec{r}_i - \vec{r}_k) +
\sum_{i<j\in\mathcal{P}_{\pi},k\in\mathcal{P}_{K}}\eta_{3,\pi\pi K}\delta^{(3)}(\vec{r}_i - \vec{r}_j)\delta^{(3)}(\vec{r}_i - \vec{r}_k) + \nonumber \\
& \sum_{i\in\mathcal{P}_{\pi},j<k\in\mathcal{P}_{K}}\eta_{3,\pi K
K}\delta^{(3)}(\vec{r}_i - \vec{r}_j)\delta^{(3)}(\vec{r}_i -
\vec{r}_k) + \sum_{i<j<k\in\mathcal{P}_{K}}\eta_{3,K K
K}\delta^{(3)}(\vec{r}_i - \vec{r}_j)\delta^{(3)}(\vec{r}_i -
\vec{r}_k)
\end{align}
where the first line is the two-body interaction potential while the
remaining lines constitute the various three-body potentials. The
sets $\mathcal{P}_{\pi}$ and $\mathcal{P}_{K}$ refer to the set of
all pions and kaons in the system, respectively. Due to the
requirement that bound states do not form, the interactions between
pions and kaons studied are restricted to be repulsive. Hence all
$\pi$ and $K$ fields referenced are shorthand for $\pi^{+}$ and
$K^{+}$. Because the particles are nonrelativistic, interactions
such as $KK\to\pi\pi$ or $\pi\pi\to{KK}$ are not considered.

The $\chi$'s can be related to the scattering amplitude and for
small external momentum be expanded using effective range theory
according to $\chi = -\frac{4\pi}{M} p^{-1} \ \text{tan} \ \delta(p)
\to -\frac{4\pi}{M}(-\frac{1}{a} + \frac{1}{2}r_0 p^2 +
\cdots)^{-1}$ where $a$ is the $s$-wave scattering length and $r_0$
is the effective range.  The modern language and description of this
approach is pionless effective field theory or EFT($\pi
\!\!\!/$)\cite{Kaplan:1998tg,vanKolck:1998bw,Chen:1999tn}. In this
framework, the two-body interaction $\chi$ corresponds to the
expansion of the $2 \to 2$ contact interaction in the
Lagrangian\cite{vanKolck:1998bw}. The language of EFT($\pi \!\!\!/$)
possesses a more natural way of dealing with issues of
renormalization and power counting as the choice of a particular
subtraction scheme leads to order-by-order renormalization of loop
divergences.

The contact potential from eqn. (\ref{potential}) can be reexpressed
in terms of a Hamiltonian in NRPT, as described in ref.
\cite{Detmold:2008gh}. Using this method, each term in the large
volume expansion of the function $r(z)$ can be expressed as a sum of
diagrams in the perturbation theory. Vertices in these diagrams are
given by the $\chi$ and $\eta$ shown above, while the propagators
are given by the expectation of the free theory resolvent
$\langle\frac{1}{z-\mathcal{H}_0}\rangle$ over intermediate states.
The presence of the projection operator $\hat{Q}_0$ removes from
consideration any diagrams where the intermediate state expectation
gives $\langle\mathcal{H}_0\rangle=0$. Loop divergences are
regulated with dimensional regularization in our calculation. The
divergences arising from two-loop diagrams will necessitate the
inclusion of three-body interaction strengths as a function of the
renormalization scale $\mu$. In finite volume the integrals normally
associated with the continuum expression for each diagram will
become sums because the momenta will be restricted to the possible
values $p=2\pi\vec{n}/L$ for $n_{j}\in \mathbb{Z}$.

It is simple to show that generating all of the diagrams and summing
their contributions leads to the following result for the ground
state energy shift:
\begin{eqnarray}
    \Delta{E}_{0}(n,m,L)&=&{E}_{0}(n,m,L)-n \ m_{\pi}-m \
    m_{K}\nonumber\\
    &=&E_{\pi}(n,L)+E_{K}(m,L)+E_{\pi K}(n,m,L)
\end{eqnarray}
with ($m_{\pi{K}}=m_{\pi}m_{K}/(m_{\pi}+m_{K})$)
\begin{eqnarray}
    E_{i}(n,L)&=&\frac{4\pi{\bar{a}_i}}{m_{i}L^3}\left(\begin{array}{c}n\\2\end{array}\right)\left[1-\left(\frac{\bar{a}_i}{\pi{L}}\right)\mathcal{I}
    +\left(\frac{\bar{a}_i}{\pi{L}}\right)^{2}\left(\mathcal{I}^2+(2n-5)\mathcal{J}\right)\right.\nonumber\\
    &&\left.-\left(\frac{\bar{a}_i}{\pi{L}}\right)^3\left(\mathcal{I}^3+(2n-7)\mathcal{I}\mathcal{J}+(5n^2-41n+63)\mathcal{K}\right)\right]
    +\left(\begin{array}{c}n\\3\end{array}\right)\frac{\bar{\eta}_{3,i}(\mu)}{L^6}
    +\mathcal{O}(L^{-7})
\end{eqnarray}
\begin{eqnarray}\label{piKanswer}
    E_{\pi
    K}(n,m,L)&=&\frac{2{\pi}\bar{a}_{\pi{K}}mn}{m_{\pi{K}}L^3}\left[1-\left(\frac{\bar{a}_{\pi{K}}}{\pi{L}}\right)\mathcal{I}\right.\nonumber\\
    &&\left.+\left(\frac{\bar{a}_{\pi{K}}}{\pi{L}}\right)^{2}\left(\mathcal{I}^2
    +\mathcal{J}\left[-1+\frac{\bar{a}_{\pi}}{\bar{a}_{\pi{K}}}(n-1)\left(\frac{1}{m_{\pi{K}}}+\frac{2}{m_{\pi}}\right)
    +\frac{\bar{a}_{K}}{\bar{a}_{\pi{K}}}(m-1)\left(\frac{1}{m_{\pi{K}}}+\frac{2}{m_{K}}\right)\right]\right)\right.\nonumber\\
    &&\left.+\left(\frac{\bar{a}_{\pi{K}}}{\pi{L}}\right)^{3}\left(-\mathcal{I}^{3}
    +f^{\mathcal{K},\pi{K}}\left(\frac{\bar{a}_{\pi}\bar{a}_{K}}{\bar{a}_{\pi{K}}^2}\right)\mathcal{K}
    +\sum_{i=0}^{2}\sum_{p=\pi,K}\left(f^{\mathcal{I}\mathcal{J},p}_{i}\mathcal{I}\mathcal{J}
    +f^{\mathcal{K},p}_{i}\mathcal{K}\right)\left(\frac{\bar{a}_{p}}{\bar{a}_{\pi{K}}}\right)^{i}\right)
    \right]\nonumber\\
    &&+\frac{nm(n-1)\bar{\eta}_{3,\pi\pi{K}}(L)}{2L^6}+\frac{nm(m-1)\bar{\eta}_{3,\pi{KK}}(L)}{2L^6}+\mathcal{O}(L^{-7})
\end{eqnarray}
with the scattering lengths given in terms of the parameters
$\bar{a}_i=\frac{M\chi_i}{4\pi}$ and the effective ranges $r_i$
by\cite{Detmold:2008fn}
\begin{eqnarray}
    a_{\pi}^{I=2}&=&\bar{a}_{\pi}-\frac{2{\pi}\bar{a}_{\pi}^{3}r_{\pi}^{I=2}}{L^3}\nonumber\\
    a_{K}^{I=1}&=&\bar{a}_{K}-\frac{2{\pi}\bar{a}_{K}^{3}r_{K}^{I=1}}{L^3}\nonumber\\
    a_{\pi{K}}^{I=3/2}&=&\bar{a}_{\pi{K}}-\frac{2{\pi}\bar{a}_{\pi{K}}^{3}r_{\pi{K}}^{I=3/2}}{L^3}.
\end{eqnarray}
The four volume dependent (but renormalization scale independent)
quantities from the three-body interactions are defined by
($y=m_{\pi}/m_{K}$):
\begin{eqnarray}
    \bar{\eta}_{3,i}(L)&=&\eta_{3,i}(\mu)+\frac{64\pi{a_i}^4}{m_i}(3\sqrt{3}-4\pi){\rm
    log}(\mu{L})-\frac{96a_i^4}{{\pi^2}m_i}(2q[1,1]+r[1,1])\nonumber\\
    \bar{\eta}_{3,\pi\pi{K}}(L,y)&=&\eta_{3,\pi\pi{K}}(\mu,y)
    -\frac{4a_{\pi{K}}^4}{{\pi^2}m_{\pi{K}}}\sum_{i=0}^{2}\sum_{p=\pi,K}\sum_{\mathcal{N}\in\mathcal{N}_1}
    \left(\frac{a_{p}}{a_{\pi{K}}}\right)^{i}f_{i}^{\mathcal{N},p}\mathcal{N}\nonumber\\
    \bar{\eta}_{3,\pi{KK}}(L,y)&=&\eta_{3,\pi{KK}}(\mu,y)
    -\frac{4a_{\pi{K}}^4}{{\pi^2}m_{\pi{K}}}\sum_{i=0}^{2}\sum_{p=\pi,K}\sum_{\mathcal{N}\in\mathcal{N}_2}
    \left(\frac{a_{p}}{a_{\pi{K}}}\right)^{i}f_{i}^{\mathcal{N},p}\mathcal{N}
\end{eqnarray}
and
\begin{eqnarray}
    \mathcal{N}_1&=&\left\{\hat{Q}(1,y),\hat{Q}(y,1),\hat{R}(y,1),\hat{R}(1/y,1/y)\right\}\nonumber\\
    \mathcal{N}_2&=&\left\{\hat{Q}(1,1/y),\hat{Q}(y,y),\hat{R}(y,y),\hat{R}(1,1/y)\right\}.
\end{eqnarray}

The functions $\hat{Q}$, $\hat{R}$, $q$, and $r$ are defined in the
appendix along with the coefficients $f_i$. The finite parts of
$\hat{Q}(a,b)$ and $\hat{R}(a,b)$ are scheme dependent quantities
where changes in the value will be absorbed by $\eta_3(\mu)$. The
numerical values for the MS scheme are given in the appendix.
However, the $\bar{\eta}_3$ are not scheme dependent and this is the
quantity that will be determined during a lattice calculation.
Furthermore, note that the three-body interactions in the
$\pi\pi{K}$ and $\pi{KK}$ cases depend on the $\pi{K}$ mass ratio.
One can immediately see that this is necessary if one takes the
limiting case where either the pion or the kaon become infinitely
heavy, where the heavy particle decouples from the theory and hence
all cross species interactions must go to zero. Finally, in the
limit of $n\to0$ or $m\to0$ this result simplifies to the previously
determined $n$-boson case while the limit $m\to n$ with $m_{K}\to
m_{\pi}$ and all interactions set to be equal it simplifies to the
$2n$-boson case. The result given agrees with previously calculated
results in the single species limit
\cite{Detmold:2008gh,Beane:2007qr,Tan:2007bg}.

In this work we have calculated the ground state energy shift of the
mixed species system of $n$ pions and $m$ kaons. Using this result a
rigorous connection can be made between Euclidean-space lattice QCD
calculations of mixed pion-kaon systems and Minkowski-space
multibody interaction strengths. In Ref. \cite{Detmold:2008yn} the
isospin and strangeness chemical potentials are analyzed with
lattice QCD at values relevant to dense nuclear matter for purely
kaon condensates. With the results presented in this paper such an
analysis can be done for mixed pion-kaon systems, allowing lattice
calculations of the $\pi{K}$ scattering length and three-body
interaction strengths $\pi\pi{K}$ and $\pi{KK}$. Such extractions
are necessary for examinations of the nuclear equation of state for
values of the isospin and strangeness chemical potentials where it
is not energetically favorable to form either pure pion or pure kaon
condensates, but where possibly mixed systems of pions and kaons
exist.

\acknowledgments B.S. and J.W. would like to thank Martin Savage and
William Detmold for their time and many useful discussions.

\appendix
\section*{Appendix}
\renewcommand{\theequation}{A-\arabic{equation}}
\setcounter{equation}{0} The coefficients $f_i$ are given by (where
$y=m_{\pi}/m_{K}$ and $m_{\pi{K}}=m_{\pi}m_{K}/(m_{\pi}+m_{K})$):
\begin{eqnarray}
    f^{\mathcal{I}\mathcal{J},\pi}_{0}&=&\frac{1}{2}\left(1+m+n\right)\\
    f^{\mathcal{I}\mathcal{J},\pi}_{1}&=&2\left(1+\frac{2m_{\pi{K}}}{m_{\pi}}\right)(1-n)\\
    f^{\mathcal{I}\mathcal{J},\pi}_{2}&=&m_{\pi{K}}\frac{m_{\pi}-m_{K}}{m_{\pi}m_{K}}(1-n)\\
    f^{\mathcal{I}\mathcal{J},K}_{i}&=&f^{\mathcal{I}\mathcal{J},\pi}_{i}(n\leftrightarrow{m},m_{\pi}\leftrightarrow{m_{K}})
\end{eqnarray}
\begin{eqnarray}
    f^{\mathcal{K},\pi}_{0}&=&\frac{1}{2}\left(4n+4m-3mn-6+\frac{m_{\pi}^2+m_{K}^2}{m_{\pi}m_{K}}(m+n-mn-1)\right)\\
    f^{\mathcal{K},\pi{K}}&=&-8(n-1)(m-1)\left(1+\frac{m_{\pi{K}}^{2}}{m_{\pi}m_{K}}\right)\\
    f^{\mathcal{K},\pi}_{1}&=&\frac{2m_{\pi{K}}(m_{\pi}^2+9m_{K}^2+4m_{\pi}m_{K})}{m_{\pi}m_{K}^2}(n-1)\\
    f^{\mathcal{K},\pi}_{2}&=&\frac{m_{\pi{K}}^2}{m_{\pi}^2m_{K}^3}(1-n)\left[m_{K}^{3}(13n-45)-m_{\pi}^{3}+m_{\pi}m_{K}^2(14n-39)+5m_{\pi}^2m_{K}(n-3)\right]\nonumber\\
    &&-\frac{m_{\pi{K}}(3m_{K}+m_{\pi})^2}{m_{\pi}^2m_{K}^2\left(\frac{1}{m_{\pi{K}}}+\frac{2}{m_{\pi}}\right)}(n-1)(n-2)\\
    f^{\mathcal{K},K}_{i}&=&f^{\mathcal{K},\pi}_{i}(n\leftrightarrow{m},m_{\pi}\leftrightarrow{m_{K}})
\end{eqnarray}
\begin{eqnarray}
    f^{\hat{\mathcal{Q}}(1,y),\pi}_{0}&=&f^{\hat{\mathcal{R}}(y,1),\pi}_{0}=\frac{m_{\pi}}{m_{\pi{K}}}\\
    f^{\hat{\mathcal{Q}}(1,y),\pi}_{1}&=&f^{\hat{\mathcal{R}}(y,1),\pi}_{1}=\frac{2m_{\pi}}{m_{\pi{K}}}\\
    f^{\hat{\mathcal{Q}}(1,y),\pi}_{2}&=&f^{\hat{\mathcal{R}}(y,1),\pi}_{2}=\frac{m_{\pi}}{m_{\pi{K}}}\\
    f^{\hat{\mathcal{Q}}(1,1/y),K}_{i}&=&f^{\hat{\mathcal{R}}(1,1/y),K}_{i}=f^{\hat{\mathcal{Q}}(1,y),\pi}_{i}(m_{\pi}\leftrightarrow{m_{K}})\\
    f^{\hat{\mathcal{Q}}(1,y),K}_{i}&=&f^{\hat{\mathcal{R}}(y,1),K}_{i}=f^{\hat{\mathcal{Q}}(1,1/y),\pi}_{i}=f^{\hat{\mathcal{R}}(1,1/y),\pi}_{i}=0
\end{eqnarray}
\begin{eqnarray}
    f^{\hat{\mathcal{Q}}(y,1),\pi}_{0}&=&f^{\hat{\mathcal{Q}}(y,1),K}_{0}=f^{\hat{\mathcal{Q}}(y,1),K}_{1}=f^{\hat{\mathcal{Q}}(y,1),K}_{2}=f^{\hat{\mathcal{Q}}(y,y),\pi}_{i}=0\\
    f^{\hat{\mathcal{Q}}(y,1),\pi}_{1}&=&8\\
    f^{\hat{\mathcal{Q}}(y,1),\pi}_{2}&=&8\\
    f^{\hat{\mathcal{Q}}(y,y),K}_{i}&=&\frac{m_{\pi}}{m_{K}}f^{\hat{\mathcal{Q}}(y,1),\pi}_{i}
\end{eqnarray}
\begin{eqnarray}
    f^{\hat{\mathcal{R}}(y,y),\pi}_{0}&=&f^{\hat{\mathcal{R}}(y,y),\pi}_{1}=f^{\hat{\mathcal{R}}(y,y),\pi}_{2}
    =f^{\hat{\mathcal{R}}(y,y),K}_{0}=f^{\hat{\mathcal{R}}(y,y),K}_{1}=f^{\hat{\mathcal{R}}(1/y,1/y),K}_{i}=0\\
    f^{\hat{\mathcal{R}}(y,y),K}_{2}&=&\frac{8m_{\pi}m_{\pi{K}}}{m_{K}^2}\\
    f^{\hat{\mathcal{R}}(1/y,1/y),\pi}_{i}&=&f^{\hat{\mathcal{R}}(y,y),K}_{i}(m_{\pi}\leftrightarrow{m_{K}})
\end{eqnarray}

The integer sums that contribute to $\mathcal{O}(L^{-6})$ include:
\begin{eqnarray}
\mathcal{I} & =&\lim_{\Lambda_{j}\to\infty}\left( \sum_{\vec{n} \neq 0}^{\Lambda_j} \frac{1}{|\vec{n}|^2} -4{\pi}{\Lambda_j}\right)= -8.9136329\\
\mathcal{J} & =& \sum_{\vec{n} \neq 0} \frac{1}{|\vec{n}|^4} = 16.532316 \\
\mathcal{K} & =& \sum_{\vec{n} \neq 0} \frac{1}{|\vec{n}|^6} = 8.4019240\\
\mathcal{Q}(a,b) & =& \sum_{\vec{n} \neq 0} \sum_{\vec{m} \neq 0} \frac{1}{|\vec{n}|^2 |\vec{m}|^2}\frac{1}{|\vec{n}|^2+a|\vec{m}|^2+b(\vec{n} + \vec{m})^2} \label{qq}\\
\mathcal{R}(a,b) & =& \sum_{\vec{n} \neq 0} \left[\sum_{\vec{m}}
\frac{1}{|\vec{n}|^4}\frac{1}{|\vec{n}|^2+a|\vec{m}|^2+b(\vec{n} +
\vec{m})^2}-\frac{1}{a+b}\int{d^{d}n}\frac{1}{|\vec{n}|^2}\right]\label{rr}
\end{eqnarray}
where the $\mathcal{Q}$ and $\mathcal{R}$ sums need to be regulated
(defined) using dimensional regulation and in the $\mathcal{R}$ sum
the subtracted integral removes a nested divergence that would
otherwise affect the one-loop scattering amplitude. One can
calculate
\begin{eqnarray}
    \mathcal{Q}(a,b)&=&\mathcal{Q}_{\Sigma}(a,b)+\mathcal{Q}_{DR}(a,b)-\mathcal{Q}_{\Lambda}(a,b)\nonumber\\
    &=&\hat{\mathcal{Q}}(a,b,\mu)+\frac{2{\pi^3}}{b}{\rm
    sin}^{-1}\left(\frac{b}{\sqrt{(a+b)(1+b)}}\right)\frac{1}{\epsilon}\\
    \mathcal{R}(a,b)&=&\mathcal{R}_{\Sigma}(a,b)+\mathcal{R}_{DR}(a,b)-\mathcal{R}_{\Lambda}(a,b)\nonumber\\
    &=&\hat{\mathcal{R}}(a,b,\mu)-\frac{2{\pi^3}\sqrt{a+b+ab}}{(a+b)^2}\frac{1}{\epsilon}
\end{eqnarray}
where the cutoff dependencies of the sums $\mathcal{Q}_{\Sigma}$ and
$\mathcal{R}_{\Sigma}$ are canceled by the cutoff dependence of the
$\mathcal{Q}_{\Lambda}$ and $\mathcal{R}_{\Lambda}$ integrals. The
dimensionally regularized integrals in turn counter the remaining
portions of the cutoff integrals, leaving one with the functions
$\hat{\mathcal{Q}}$ and $\hat{\mathcal{R}}$ found in eqn.
(\ref{piKanswer}). These are given by
\begin{eqnarray}
    \hat{\mathcal{Q}}(a,b,\mu)&=&q(a,b)+\frac{8{\pi^3}}{b}{\rm
    sin}^{-1}\left(\frac{b}{\sqrt{(a+b)(1+b)}}\right){\rm
    log}(\mu{L})\\
    \hat{\mathcal{R}}(a,b,\mu)&=&r(a,b)-\frac{8{\pi^3}\sqrt{a+b+ab}}{(a+b)^2}{\rm
    log}(\mu{L})
\end{eqnarray}
where in Tables \ref{table:q} and \ref{table:r} the numerical
portions of $q(a,b)$ and $r(a,b)$ are explicitly evaluated for
values of the arguments $a$ and $b$ relevant to the calculation.
\begin{table}[h]
\caption{Numerical evaluation of the function q(a,b).} 
\centering      
\begin{tabular}{| c | c c c c |}  
\hline                        

$y=\frac{m_{\pi}}{m_K}$ & $q(1,y)$ & $q(1,1/y)$ & $q(y,1)$ & $q(y,y)$ \\  [2.0ex] 
\hline                    
0.2 & -167.96763 & -34.22227 & -94.42134 & -113.71723  \\   
0.2827 & -157.34173 & -45.55930 & -95.75651 & -127.15818 \\
0.4 & -144.45515 & -59.06850 & -97.36636 & -129.53719 \\
0.6 & -126.85940 & -77.19563 & -99.49508 & -121.84065 \\
0.8 & -113.15457 & -91.10749 & -101.03837 & -111.68392 \\
1 & -102.15598 & -102.15598 & -102.15598 & -102.15598 \\ [2ex]      
\hline     
\end{tabular}
\label{table:q}  
\end{table}
\begin{table}[h]
\caption{Numerical evaluation of the function r(a,b).} 
\centering      
\begin{tabular}{| c | c c c c |}  
\hline                        

$y=\frac{m_{\pi}}{m_K}$ & $r(y,1)$ & $r(1,1/y)$ & $r(1/y,1/y)$ & $r(y,y)$ \\  [2.0ex] 
\hline                    
0.2 & 49.695247 & -0.358353 & -0.141373 & 517.146558  \\   
0.2827 & 44.958613 & 0.993223 & 0.391640 & 265.289351 \\
0.4 & 38.951624 & 3.803394 & 1.738871 & 133.442877 \\
0.6 & 30.564875 & 9.279473 & 5.616888 & 58.127093  \\
0.8 & 24.132625 & 14.530094 & 11.449288 & 31.463511 \\
1 & 19.186903 & 19.186903 & 19.186903 & 19.186903 \\ [2ex]      
\hline     
\end{tabular}
\label{table:r}  
\end{table}

\bibliography{masterbibtex}

\end{document}